\documentclass{article}
\usepackage[utf8]{inputenc}
\usepackage[backend=biber, style=numeric]{biblatex}
\usepackage{hyperref}
\usepackage{xurl}
\hypersetup{breaklinks=true}
\usepackage{amsfonts}
\usepackage{amsthm}
\usepackage{amsmath}
\usepackage{setspace}
\usepackage{verbatim}
\usepackage{tikz-cd}
\usepackage{enumitem}
\usepackage{tensor}
\usepackage{pifont}
\usepackage{enumitem}
\usepackage{txfonts}
\usepackage{mathtools}
\usepackage{stmaryrd}
\usepackage[T1]{fontenc}
\usepackage[a4paper, total={6in, 8in}]{geometry}

\title{A Supergeometric Fa\`{a} di Bruno Formula}
\author{Andreas Swerdlow \\ \small The University of Manchester, UK \\ \small\texttt{andreas.swerdlow@manchester.ac.uk}}
\date{August 2024}

\newtheorem{theorem}{Theorem}[section]

\newtheorem{proposition}[theorem]{Proposition}

\newtheorem{definition}[theorem]{Definition}

\newtheorem*{declarations}{Declarations}
\newtheorem*{acknowledgement}{Acknowledgement}

\newtheorem*{faa_di_bruno}{Fa\`{a} di Bruno's Formula}
\newtheorem*{super_faa_di_bruno}{Super Fa\`{a} di Bruno's Formula}

\newcommand{\R}{\mathbb{R}}
\newcommand{\Z}{\mathbb{Z}}

\newcommand{\dely}[1]{\frac{\partial}{\partial y^{#1}}}
\newcommand{\delx}[1]{\frac{\partial}{\partial x^{#1}}}
\newcommand{\delyof}[2]{\frac{\partial #1}{\partial y^{#2}}}
\newcommand{\delxof}[2]{\frac{\partial #1}{\partial x^{#2}}}

\newcommand{\delxi}[1]{\frac{\partial}{\partial \xi^{#1}}}
\newcommand{\delxiof}[2]{\frac{\partial #1}{\partial \xi^{#2}}}

\DeclareMathOperator{\partition}{Part}
\newcommand{\superspace}[2]{\R^{#1\,|\,#2}}
\newcommand{\superalg}[2]{C^\infty(\R^{#1\,|\,#2})}

\tikzstyle{vertex}=[circle, draw, inner sep=0pt, minimum size=6pt]

\begin{document}

\maketitle

\begin{abstract}
    We extend the multivariate Fa\`{a} di Bruno formula to the super case, where anticommuting odd coordinates are considered. The formula takes the same form as the classical case but contains some nontrivial signs, which essentially measure the failure to order the odd factors and derivatives optimally. As a quick application, we obtain an explicit combinatorial formula for the generalized super Bell polynomials, defined by Fan and Hon.
\end{abstract}

\section{Introduction}\label{sec_intro}

In basic single variable calculus, Fa\`{a} di Bruno's formula \footnote{Francesco Fa\`{a} di Bruno's papers stating and proving the formula are predated by a number of works (for details of the history see \cite{FdB_history}, \cite{FdB_prehistory}), the earliest being Louis François Antoine Arbogast's Trait\'e du
Calcul des D\'erivations.} is to the chain rule what the more well known Leibniz rule is to the product rule. That is, it gives an explicit way to calculate multiple derivatives of a composite function. There are a number of ways to write the formula, though we will focus only on the so called combinatorial form, which is written as a sum over partitions. For any set $A$, we denote by $\partition(A)$ the set of partitions of $A$. Then the combinatorial form of Fa\`{a} di Bruno's formula is:
\begin{faa_di_bruno}
    \begin{equation}
        \frac{d^n}{dx^n} f(y(x)) = \sum_{\pi \in \partition(\{1, \dots, n\})} f^{(|\pi|)}(y(x)) \prod_{B \in \pi} \frac{d^{|B|}y(x)}{dx^{|B|}} .
    \end{equation}
\end{faa_di_bruno}

Multivariable versions of the formula can be found in at least two papers, by Constantine and Savits \cite{Constantine_Savits}, and Leipnik and Pearce \cite{Leipnik_Pearce}. However, in \cite{Hardy_combinatorics}, Hardy gives a version which is (in our opinion) conceptually simpler, elucidates the combinatorial aspects of the formula, and more directly applies to situations where the partial derivatives are taken with respect to some, possibly unspecified, collection of indices. This last situation occurs all the time, especially in the author's field of differential geometry, for example when one is applying a differential operator written in the form $S^{a_1 \dots a_n}(x)\delx{a_1}\cdots\delx{a_n}$, where the contracted indices $a_1, \dots, a_n$ are meant to be summed over by the Einstein summation convention, and $S^{a_1 \dots a_n}(x)$ is some rank $n$ contravariant symmetric tensor.

We will now state Hardy's version of the formula. Let $x = (x^1, \dots, x^d)$ be some set of suitable variables (e.g. coordinates on a smooth manifold), and let $a_1, \dots, a_n \in \{1, \dots, d\}$ be some set of indices. Let $y(x) = y(x^1, \dots , x^d)$ be some real valued function (with enough derivatives) depending on $x$, and $f : \R \rightarrow \R$ another function also with enough derivatives.

\begin{equation}
    \delx{a_n} \dots \delx{a_1} f(y(x)) = \sum_{\pi \in \partition(\{a_1, \dots, a_n\})} f^{(|\pi|)}(y(x)) \prod_{B \in \pi} \frac{\partial^{|B|}y}{\prod_{j \in B} \partial x^{a_j}} 
\end{equation}

We can make this slightly more general by allowing $y$ to be a vector valued function $y(x) \in \R^k$ and letting $f: \R^k \rightarrow \R$. The formula then becomes
\begin{equation}
    \delx{a_n} \dots \delx{a_1} f(y(x)) = \sum_{\pi \in \partition(\{a_1, \dots, a_n\})}   \frac{\partial^{|\pi|} f(y(x))}{\partial^{b_1} \cdots \partial^{b_{|\pi|}}} \prod_{i = 1}^{|\pi|}  \frac{\partial^{|B^i_\pi|}y^{b_i}(x)}{\prod_{j \in B^i_\pi} \partial x^{a_j}} .
\end{equation}

The purpose of this paper is to extend Fa\`{a} di Bruno's formula to the super case. In supergeometry one extends the usual commuting "even" coordinates (such as coordinates $x^i$ on a smooth manifold), by adding "odd" coordinates, which anticommute with each other. The prefix "super" comes from theoretical physics, where supergeometry is the geometry underlying supersymmetric field theories. These field theories have two types of field, the commutative "bosonic" fields and the anticommutative "fermionic" fields, and so even coordinates in supergeometry are sometimes called bosonic coordinates, and odd coordinates are called fermionic coordinates. For introductory texts on supergeometry see \cite{Berezin2010-ay}, \cite{Voronov_2014}, \cite{Leites1980-jz}, and \cite{Cattaneo_Schaetz}.

To quote Th. Voronov in \cite{Voronov_2014}, section 2.1.4: "In general all naturally formulated analogues of the assertions in an analysis course carry over to the supercase." In particular, any identities or theorems on differentiation extend naturally, as long as one is careful with the ordering of factors and possibly with inclusions of extra signs, due to the presence of odd variables. We will show that Fa\`{a} di Bruno's formula is no different, although due to the comparative complexity of the formula, the ordering and signs will be fairly complicated (and constitutes the only original contribution of this article).   

The classical Fa\`{a} di Bruno formula is intimately related to the Bell polynomials, as each can be written in terms of the other. In section \ref{sec_super_Bell}, we show that this story extends to the super case, where the classical Bell polynomials are replaced by the generalized super Bell polynomials, as defined by Fan and Hon in \cite{super_Bell}. These multivariable differential polynomials are shown in \cite{super_Bell} to be useful for studying infinite conservation laws of supersymmetric equations, such as the supersymmetric KdV equation and supersymmetric sine-Gordon equation.

\begin{acknowledgement}
    I'd like to thank my supervisor Ted Voronov for his encouragement and guidance in writing this paper. 
\end{acknowledgement}

\section{Basic Differential Calculus on Superspaces}

\subsection{Functions of Odd Variables}

We are interested in smooth functions depending on even and odd variables. By even variables we mean coordinates $x^i$ on $\R^n$ or more generally any $n$-manifold $M$. By odd variables we mean generators $\xi^\mu$ of the $m$-dimensional Grassmann algebra, which is just the free algebra generated by $\xi^\mu$ modulo the anticommutation relation 
\begin{equation}\label{eq_Grassmann_relation}
    \xi^\mu\xi^\nu = - \xi^\nu\xi^\mu.
\end{equation}
Equation \eqref{eq_Grassmann_relation} implies that the $\xi^\mu$ are all nilpotent, so any smooth function depending on the $\xi^\mu$ is constrained to be a polynomial of maximum degree $m$. So a smooth a function $f(x,\xi)$ depending on $x^i$ and $\xi^\mu$ should take the form
\begin{equation}\label{eq_function_genform}
    f(x,\xi) = f_0(x) + \xi^\mu f_\mu(x) + \xi^{\mu_1}\xi^{\mu_2} f_{\mu_1\mu_2}(x) + \dots + \xi^{\mu_1}\dots\xi^{\mu_m} f_{\mu_1 \dots \mu_m}(x).
\end{equation}
The coefficients are all smooth functions and antisymmetric in their indices. Functions of the form \eqref{eq_function_genform}, if we take the $x^i$ to be coordinates on $\R^n$, form an algebra which we denote by $\superalg{n}{m}$. We think of this as the algebra of smooth functions on the superspace $\superspace{n}{m}$, where $n$ is the even dimension and $m$ the odd dimension. We call $(x^1 , \dots, x^n; \,\xi^1, \dots, \xi^m)$ collectively a set of coordinates on $\superspace{n}{m}$. Note that there is no actual space (by which we mean a set with some geometrical/topological structure) for which $\superalg{n}{m}$ is the algebra of functions. 

$\superalg{n}{m}$ is a $\Z_2$-graded algebra, that is $\superalg{n}{m} = C^\infty_0(\superspace{n}{m}) \oplus C^\infty_1(\superspace{n}{m})$, where $C^\infty_0(\superspace{n}{m})$ consists of functions of the form \eqref{eq_function_genform} with $f_{\mu_1 \dots \mu_k}(x) = 0$ when $k$ is odd, and $C^\infty_1(\superspace{n}{m})$ consists of functions with $f_{\mu_1 \dots \mu_k}(x) = 0$ when $k$ is even. We say elements of $C^\infty_0(\superspace{n}{m})$ are even/have parity 0, and and elements of $C^\infty_1(\superspace{n}{m})$ are odd/have parity 1. For a "homogeneous" element $f(x,\xi)$ of either $C^\infty_0(\superspace{n}{m})$ or $C^\infty_1(\superspace{n}{m})$, we denote by $\tilde{f}$ its parity. The algebra $\superalg{n}{m}$ is supercommutative, i.e. for any homogeneous functions $f, g$ we have
\begin{equation}
    fg = (-1)^{\tilde{f}\tilde{g}} gf.
\end{equation}
So even functions commute with everything and odd functions anticommute with each other.

\subsection{Substitution/Change of Variables for Functions of Odd Variables}

The Fa\`{a} di Bruno formula involves the derivative of a function in a new set of variables, so we now describe how substition/change of variables works in the super setting. Given the superspaces $\superspace{n_1}{m_1}$ and $\superspace{n_2}{m_2}$ with respective coordinates $(x,\xi)$ and $(y,\zeta)$, a morphism/smooth map $\superspace{n_1}{m_1} \rightarrow \superspace{n_2}{m_2}$ is defined by $n_2$ even functions
\begin{equation*}
    y = (y^1(x,\xi), \dots, y^{n_2}(x,\xi)) \in \superalg{n_1}{n_2}^{\times n_2} ,
\end{equation*}
and $m_2$ odd functions 
\begin{equation*}
    \zeta = (\zeta^1(x,\xi), \dots, \zeta^{m_2}(x,\xi)) \in \superalg{n_1}{n_2}^{\times m_2} .
\end{equation*}
Then, given a function $f(y,\zeta) = f_0(y) + \zeta^\alpha f_\alpha(y) + \zeta^{\alpha_1}\zeta^{\alpha_2} f_{\alpha_1\alpha_2}(y) + \dots \in \superalg{n_2}{m_2}$, we want to define the substitution $f(y(x,\xi),\zeta(x,\xi)) \in \superalg{n_1}{m_1}$ (also called the pullback by the smooth map), which should be a function of $x$ and $\xi$. The problem is that the $y^b(x,\xi)$ are allowed to contain terms with an even number of factors of $\xi^\mu$'s, and we don't a priori know how to evaluate the coefficients $f_{\alpha_1 \alpha_2\dots}(y)$ on these terms (since the coefficients should just be smooth functions on $\R^{n_2}$). To solve this problem, we use a procedure called Grassmannian analytic continuation, which uniquely extends every smooth function $\R^{n} \rightarrow \R$ to a function $\R^{n} \otimes \Lambda(\xi^1, \dots, \xi^m) \rightarrow \R \otimes \Lambda(\xi^1, \dots, \xi^m)$. The procedure in our case is:
\begin{enumerate}
    \item Each $y^b(x,\xi)$ can be written as $y^b(x,\xi) = y^b_0 + y^b_+$, where $y^b_0 := y^b(x,0) \in \R$ is the "numerical" part containing no factors of $\xi^\mu$'s, and $y^b_+ := y^b - y^b_0$ is the "nilpotent supplement". We write $y_0 = (y^1_0, \dots, y^{n_2}_0)$ and $y_+ = (y^1_+, \dots, y^{n_2}_+)$ so $y(x,\xi) = y_0 + y_+$
    \item For each coefficient $f_{\alpha_1 \alpha_2\dots}(y)$ of $f$ we perform a Taylor expansion about $y_0$
    \begin{equation*}
        f_{\alpha_1 \alpha_2\dots}(y_0 + \epsilon) = f_{\alpha_1 \alpha_2\dots}(y_0) + \epsilon^b\dely{b}f_{\alpha_1 \alpha_2\dots}(y_0) + \frac{1}{2} \epsilon^{b_2} \epsilon^{b_1}\dely{b_1}\dely{b_2}f_{\alpha_1 \alpha_2\dots}(y_0) + \dots .
    \end{equation*}
    \item The we formally make the substitution $\epsilon = y_+(x,\xi)$ to get a power series
    \begin{equation*}
        f_{\alpha_1 \alpha_2\dots}(y(x,\xi)) = f_{\alpha_1 \alpha_2\dots}(y_0 + y_+) = f_{\alpha_1 \alpha_2\dots}(y_0) + y_+^b\dely{b}f_{\alpha_1 \alpha_2\dots}(y_0) + \frac{1}{2}y^{b_2}_+y^{b_1}_+\dely{b_1}\dely{b_2}f_{\alpha_1 \alpha_2\dots}(y_0) + \dots .
    \end{equation*}
    Due to the nilpotency of each $y^b_+$, this series is finite and gives a well defined value in $\R \otimes \Lambda(\xi^1, \dots, \xi^{m_1})$, and so we get well defined function depending on $x, \xi$ for the full substitution $f(y(x,\xi),\zeta(x,\xi)) \in \superalg{n_1}{m_1}$.
 \end{enumerate}
The pullback by a smooth map $\superspace{n_1}{m_1} \rightarrow \superspace{n_2}{m_2}$ between superspaces is, just like in the classical case, an ordinary algebra homomorphism. It also clearly preserves parity of functions, so in fact it is an (even) superalgebra homomorphism.

\subsection{Differentiation}

The usual partial derivatives of even variables $\delx{a}$ can be extended to $\superalg{n}{m}$ by simply differentiating the coefficients in \eqref{eq_function_genform}, in particular we require $\delxof{\xi^\mu}{a} = 0$. Partial derivatives with respect to odd variables $\delxi{\mu}$ are completely defined by the Linearity, the product rule, and setting $\delxiof{\xi^\nu}{\mu} = \delta^\nu_\mu$.

From now on we use collective notation, where we collect the odd coordinate $\xi^\mu$ into the supercoordinates $x^a$, allowing each index $a$ to contain its parity, denoted by $\tilde{a}$, and its numerical value. Concretely, if $\tilde{a} = 0$ then $a$ can take values in $1, \dots, n$ and the new $x^a$ denotes the old even coordinate $x^a$ as above. If $\tilde{a} = 1$ then $a$ can take values in $1, \dots, m$ and the new $x^a$ denotes the odd coordinate $\xi^a$.  

Partial derivatives of functions of odd variables satisfy the usual properties of partial derivatives with two caveats: additional signs may be introduced by derivatives with respect to odd variables, and care must be taken with the ordering of factors and derivatives. Explicitly, using the collective notation described above, we have: 
\begin{align*}
    &\delx{a}(cf) = (-1)^{\tilde{a}\tilde{c}}c\delxof{f}{a}, \quad \delx{a}(f+g) = \delxof{f}{a} + \delxof{g}{a} \tag{Linearity} \\ 
    &\delx{a}(fg) = \delxof{f}{a} g + (-1)^{\tilde{a}\tilde{f}}f\delxof{g}{a} \tag{Product Rule} \\
    &\delx{a}(f(y(x)))  = \delxof{y^b}{a} \delyof{f}{b} \tag{Chain Rule} \\
    &\delx{a}\delx{b} f = (-1)^{\tilde{a}\tilde{b}}\delx{b}\delx{a} f \tag{Supercommutativity}
\end{align*}
Note that the product rule in this form requires that $f$ be homogeneous; the rule for non-homogeneous functions is found easily by splitting $f$ into its even and odd parts.

\section{The Formula}

It is clear that a super version of Fa\`{a} di Bruno's formula should take the same form as the classical formula, except with some additional signs and a specific ordering of the factors and derivatives. These signs and ordering are taken care of by the following definitions.

\begin{definition}\label{def_partition_ordering}
    Let $\pi$ be a partition of the ordered set of indices $\{ a_n, \dots, a_1 \}$. Then each block inherits the ordering of the whole set, and we order the blocks in $\pi$ by their last element, denoting them in order by $B^1_\pi, \dots, B^{|\pi|}_\pi$. The sum of the ordering on blocks of $\pi$ with the ordering within blocks also defines a new ordering on the whole set of indices, and we denote this ranking by $\pi(a_i)$ for any $i = 1, \dots, n$. For example, let $n=5$ and take the partition $ \pi = \{ \{a_1,a_4\},\{a_2,a_5\},\{a_3\}\}$. Then we have $B^1_\pi = \{a_3\}$, $B^2_\pi = \{a_5, a_2\}$, and $B^3_\pi = \{a_4, a_1\}$, and the ranking on indices is given by $\pi(a_3) < \pi(a_5) < \pi(a_2) < \pi(a_4) < \pi(a_1)$. 
\end{definition}
\begin{definition}\label{def_partition_parity}
    We define the parity $\tilde{\pi}$ of any partition $\pi$ of $\{ a_n, \dots, a_1 \}$ by
    \begin{equation}
        \tilde{\pi} \coloneqq \sum_{\substack{1 \le i,j \le n \\ i<j \,\land\, \pi(i) < \pi(j)}} \tilde{a_i} \tilde{a_j} \mod 2.
    \end{equation}
    The value of $\tilde{\pi}$ essentially measures the parity of the distance of the ordering of the partition from the original ordering of the set of indices, where only two disordered indices both having odd parity contribute. We can think of this as a special version of the Kendall tau distance between the original ordering and the ordering induced by $\pi$ on the set of indices.
\end{definition}
We are now ready to state the formula.
\begin{super_faa_di_bruno}
    \begin{equation}\label{eq_super_faa_di_bruno}
        \delx{a_n} \dots \delx{a_1} f(y(x)) = \sum_{\pi \in \partition(\{a_1, \dots, a_n\})} (-1)^{\tilde{\pi}}  \left(\prod_{i = 1}^{|\pi|}  (-1)^{\tilde{b}_i \cdot (\sum_{k=i+1}^{|\pi|}( \tilde{b}_k + \sum_{\ell \in B^k_\pi} \tilde{a}_\ell ) )} \frac{\partial^{|B^i_\pi|}y^{b_i}(x)}{\prod_{j \in B^i_\pi} \partial x^{a_j}} \right) \frac{\partial^{|\pi|} f(y(x))}{\partial^{b_1} \cdots \partial^{b_{|\pi|}}}
    \end{equation}
    Here the products $\prod_{j \in B^i_\pi} \partial x^{a_j}$ of derivatives are ordered by the ordering within the blocks, as described in definition \ref{def_partition_ordering}.
\end{super_faa_di_bruno} 
Our proof follows the same strategy as Hardy in \cite{Hardy_combinatorics}, while keeping track of signs. It relies on the standard observation that any partition of $\{1, \dots, n+1\}$ can be made from a partition of $\{1, \dots, n\}$, by either adding the element $n+1$ to an existing block, or by creating the new single element block $\{n+1\}$. 

\begin{proof}[Proof of the Formula]
    We proceed by induction on $n$. The case $n=1$ is simply the chain rule. Now assume the case $n$ and calculate
    \begin{gather*}
        \delx{a_{n+1}} \dots \delx{a_1} f(y(x)) = \delx{a_{n+1}} \left( \delx{a_n }\dots \delx{a_1} f(y(x)) \right)
        \\ = \sum_{\pi} \delx{a_{n+1}} \left[  (-1)^{\tilde{\pi}} \left(\prod_{i = 1}^{|\pi|}  \frac{\pm\partial^{|B^i_\pi|}y^{b_i}}{\prod_{j \in B^i_\pi} \partial x^{a_j}} \right) \frac{\partial^{|\pi|} f}{\partial^{b_1} \cdots \partial^{b_{|\pi|}}}  \right] \\
        = \sum_{\pi} (-1)^{\tilde{\pi}} \left[ \sum_{i=1}^{|\pi|} (-1)^{\tilde{a}_{n+1}\left( \tilde{b}_1 + \dots + \tilde{b}_{i-1} + \sum_{k=1}^{i-1}\sum_{l \in B^k_\pi} \tilde{a}_l \right)} \left( \left( \frac{\pm\partial^{|B^1_\pi|}y^{b_1}}{\prod_{j \in B^1_\pi} \partial x^{a_j}}\right) \dots  \left( \delx{a_{n+1}}\frac{\pm\partial^{|B^i_\pi|}y^{b_i}}{\prod_{j \in B^i_\pi} \partial x^{a_j}}\right) \dots \left(  \frac{\partial^{|B^{|\pi|}_\pi|}y^{b_{|\pi|}}}{\prod_{j \in B^{|\pi|}_\pi} \partial x^{a_j}}\right) \right)\frac{\partial^{|\pi|} f}{\partial^{b_1} \cdots \partial^{b_{|\pi|}}} \right. \\ \left. + (-1)^{\tilde{a}_{n+1}(\tilde{a}_n + \dots + \tilde{a}_1 + \tilde{b}_1 + \dots + \tilde{b}_{|\pi|})}\left(\prod_{i = 1}^{|\pi|} \frac{\pm\partial^{|B^i_\pi|}y^{b_i}}{\prod_{j \in B^i_\pi} \partial x^{a_j}} \right) \delxof{y^{\beta}}{a_{n+1}} \frac{\partial^{|\pi|+1} f}{\partial^{\beta} \partial^{b_1} \cdots \partial^{b_{|\pi|}}} \right] .
    \end{gather*}
    The $\pm$ signs here denote the internal signs for each block. Inside the large square brackets there are two summands. The first summand covers all partitions of $\{a_{n+1}, \dots, a_1\}$ made from $\pi$ by adding the element $a_{n+1}$ to a block. The parities of these new partitions are exactly given by the sum of $\tilde{\pi}$ and $\tilde{a}_{n+1} \sum_{k=1}^{i-1}\sum_{l \in B^k_\pi} \tilde{a}_l$ (where $i$ is the number of the block to which $a_{n+1}$ is added), and the correct block ordering does not change under this operation since it depends on the ranking of the last element of a block. The addition of $\tilde{a}_{n+1}(\tilde{b}_1 + \dots + \tilde{b}_{i-1})$ to the existing internal sign gives the correct internal sign. The second summand covers the unique partition of $\{a_{n+1}, \dots, a_1\}$ made by creating the new single element block $\{a_{n+1}\}$. Moving the term $\delxof{y^{\beta}}{a_{n+1}}$ all the way to the left of this summand, to get the correct ordering of blocks in the new partition, cancels the sign $(-1)^{\tilde{a}_{n+1}(\tilde{a}_n + \dots + \tilde{a}_1 + \tilde{b}_1 + \dots + \tilde{b}_{|\pi|})}$ and gives the new sign $(-1)^{\tilde{\beta}(\tilde{a}_n \dots \tilde{a}_1 + \tilde{b}_1 + \dots + \tilde{b}_{|\pi|})}$, which is the correct internal sign. The parity of this new partition is the same as $\tilde{\pi}$. So by the standard observation, we see that we have made terms corresponding to all the partitions of $\{a_{n+1}, \dots, a_1\}$, and we are done. 
\end{proof}

\section{Application: Generalized Super Bell Polynomials}\label{sec_super_Bell}

In \cite{super_Bell}, Fan and Hon introduce a new super version of the multivariate Bell polynomials. They call this collection of multivariable differential polynomials with respect to some smooth input function of odd variables $f(x,\xi)$, the generalized super Bell polynomials. We give their definition now. 

\begin{definition}
    For $f(x,\xi) \in C^\infty_0(\superspace{d_0}{d_1})$, an even function of odd variables, the generalized super Bell polynomials are parametrized by $\ell \in \Z^{d_0}$, and $r \in \Z_2^{d_1}$, and defined by the formula 
    \begin{equation}\label{eq_superBell_def}
        Y_{\ell,r}(f) \coloneqq e^{-f} \frac{\partial^{\ell_1}}{\partial^{\ell_1} x^{1}} \cdots \frac{\partial^{\ell_{d_0}}}{\partial^{\ell_{d_0}} x^{d_0}} \frac{\partial^{r_1}}{\partial^{r_1} \xi^{1}} \cdots \frac{\partial^{r_{d_1}}}{\partial^{r_{d_1}} \xi^{d_1}} e^{f} .
    \end{equation}
\end{definition}

To make equation \eqref{eq_superBell_def} amenable to our version of the super Fa\`{a} di Bruno formula, we re-parametrize the generalized super Bell polynomials for any ordered set of indices $\{ a_n, \dots, a_1 \}$ by the formula 
\begin{equation}
    Y_{a_1 \dots a_n}(f) \coloneqq e^{-f} \delx{a_n} \cdots \delx{a_1} e^{f} .
\end{equation}
Here, as before, $x^{a_i}$ denote collective super coordinates on $\superspace{d_0}{d_1}$. We then get the following proposition as an almost direct consequence of equation \eqref{eq_super_faa_di_bruno}.

\begin{proposition}
    For any $f(x) \in C^\infty_0(\superspace{d_0}{d_1})$, the generalized super Bell polynomials can be written in the explicit combinatorial form
    \begin{equation*}
        Y_{a_1, \dots, a_n}(f) = \sum_{\pi} (-1)^{\tilde{\pi}}  \left(\prod_{i = 1}^{|\pi|} \frac{\partial^{|B^i_\pi|}f(x)}{\prod_{j \in B^i_\pi} \partial x^{a_j}} \right) ,
    \end{equation*}
    where the products $\prod_{j \in B^i_\pi} \partial x^{a_j}$ of derivatives are ordered by the ordering within the blocks, as described in definition \ref{def_partition_ordering}, and the parity $\tilde{\pi}$ is defined by Definition \ref{def_partition_parity}. 
\end{proposition}

\begin{declarations}
    This work was supported by the Additional Funding Programme for Mathematical Sciences, delivered by EPSRC (EP/V521917/1) and the Heilbronn Institute for Mathematical Research. 
\end{declarations}
\printbibliography

\end{document}